\documentclass[11pt,a4paper]{article}
\usepackage{fullpage}
\usepackage{amsthm}
\usepackage{amsmath}
\usepackage{amssymb}
\usepackage{wrapfig} % to allow text to wrap around figures
\usepackage{url,hyperref}
\usepackage{array}
\usepackage{color}
\usepackage{float}
\usepackage[noadjust]{cite}
\usepackage{graphicx}
\usepackage{multirow}
\usepackage{ulem}
\usepackage{paralist}
\usepackage{listings}
\usepackage{bm}
\usepackage{epstopdf}

\makeatletter
\lst@Key{countblanklines}{true}[t]%
{\lstKV@SetIf{#1}\lst@ifcountblanklines}

\lst@AddToHook{OnEmptyLine}{%
	\lst@ifnumberblanklines\else%
	\lst@ifcountblanklines\else%
	\advance\c@lstnumber-\@ne\relax%
	\fi%
	\fi}
\makeatother

\long\def\ignore#1{}

\newcommand{\n}{\newline}

\newtheorem{theorem}{Theorem}[section]

\title{Efficient Partial Snapshot Implementations} %TODO Please add

\date{}

\author{Nikolaos D. Kallimanis\\
	\small{Institute of Computer Science - Foundation for Research and Technology-Hellas (FORTH-ICS)}\\
	\textit{nkallima@ics.forth.gr}
	\and
	Eleni Kanellou\\
	\small{Institute of Computer Science - Foundation for Research and Technology-Hellas (FORTH-ICS)}\\
	\textit{kanelou@ics.forth.gr}
	\and
	Charidimos Kiosterakis\\
	\small{Department of Computer Science, University of Crete, Greece}\\
	\textit{charkio@ics.forth.gr}}

\begin{document}

\maketitle
\begin{abstract}
A snapshot object is a concurrent data structure that has numerous applications in concurrent programming. Snapshots can be used to record the state of the system, so they can provide solutions to problems where an action should be taken when the global state of the system satisfies some conditions. A snapshot object consists of $m$ components, each storing a value from a given set. Processes can read/modify the state of the object by performing $UPDATE$ and $SCAN$ operations. An $UPDATE$ operation gives processes the ability to change the value of a component, while the $SCAN$ operation returns a ``consistent'' view of all the components. In most literature, two variants (in terms of the number active scanners) of  snapshot objects are studied. The first one is the single-scanner snapshot object, where at most one $SCAN$ operation is performed at any given time (whilst supporting many concurrent $UPDATE$ operations). The second one is the multi-scanner snapshot object that can support multi concurrent $SCAN$ operations at any given time. 
 
In this work, we propose the $\lambda$-scanner snapshot, a variation of the snapshot object, which supports any fixed amount of $0 < \lambda \leq n$ different $SCAN$ operations being active at any given time. 
Whenever $\lambda$ is equal to the number of processes $n$ in the system, the $\lambda$-scanner object implements a multi-scanner object, while in case that $\lambda$ is equal to $1$, the $\lambda$-scanner object implements a single-scanner object. We present the $\lambda-Snap$ snapshot object, a wait-free $\lambda$-scanner snapshot implementation that has a step complexity of $O(\lambda)$ for $UPDATE$ operations and $O(\lambda m)$ for $SCAN$ operations. The space complexity of $\lambda-Snap$ is $O(\lambda m)$. $\lambda-Snap$ provides a trade-off between the step/space complexity and the maximum number of $SCAN$ operations that the system can afford to be active on any given point in time. The low space complexity that our implementations provide makes them more appealing in real system applications.
Moreover, we provide a slightly modified version of the $\lambda-Snap$ implementation, which is called partial $\lambda-Snap$, that is able to support dynamic partial scan operations. In such an object, processes can execute modified $SCAN$ operations called $PARTIAL\_SCAN$ that could obtain a part of the snapshot object avoiding to read the whole set of components.

In this work, we first provide a simple single-scanner version of $\lambda-Snap$, which is called $1-Snap$. We provide $1-Snap$ just for presentation purposes, since it is simpler than $\lambda-Snap$. The $UPDATE$ in $1-Snap$ has a step complexity of $O(1)$, while the $SCAN$ has a step complexity of $O(m)$. This implementation uses $O(m)$ $CAS$ registers.
\end{abstract}

\section{Introduction }
\label{intro}
%
%In the last decades, \lena{we can observe the inarguably expansive applications of computer science}{applications of computer science have uarguably expanded, as} almost every device is becoming $smart$. Thus, almost every device requires to have an embedded multicore $CPU$. All this new equipment promises to solve more and more problems while performing increasingly complex jobs. In our days, more than ever, an application that does not use the many cores that are provided by the hardware is gradually becoming obsolete.
%
%A job that is performed in $t$ seconds executed by a single process can be performed in $t/n$ seconds if $n$ processes concurrently try to complete the same job. Although this is not quite true, since a complex job may have some parts that can only be executed sequentially. If we want to be more precise, a complex job may not be fully parallelizable. The portion of it that can be fully parallelized is denoted by $p$, thus, $1-p$ is the portion that has to be executed by a single core. The portion of the complex job that has to be executed by a single process dictates the maximum speedup that we can get. More specifically $Amdhal's\ Law$ \cite{Amdahl1967} implies that the best speedup that can be achieved when using infinitely many concurrent processes is $\frac{1}{1-p}$.

We inarguably live in an era where almost any activity is supported either 
by smart devices or potent servers, relying on {\it multi-core CPUs}. As this 
new equipment promises to perform more services per time unit, executing  
increasingly complex jobs, any application that does not use the many cores 
that are provided by the hardware is gradually becoming obsolete.

 At the heart of exploiting the potential that multiple cores provide, are concurrent data structures, 
 since they are essential building blocks of concurrent algorithms. The design of concurrent data structures, 
 such as lists~\cite{Timnat2012}, queues~\cite{Jayanti2005a,Kogan2011},  stacks~\cite{Bar-Nissan2011,Kogan2011}, 
 and even trees~\cite{Brown2014,Fatourou2014} is a thoroughly explored topic. Compared to sequential data structures, 
 the concurrent ones can simultaneously be accessed and/or modified by more than one process. Ideally, we would like 
 to have the best concurrent implementation, in terms of space and step complexity, of any given data structure. 
 However, this cannot always be the case since the design of those data structures is a complex task. 
 
 In this work, we present a {\it snapshot object}, a concurrent object that consists of components 
 which can be read and modified by any process. Concurrent snapshot objects are used in numerous 
 applications in order to provide a coherent ``view'' of the memory of a system. They are also used 
 to design and validate various concurrent algorithms such as the construction of concurrent timestamps~\cite{Gawlick1992}, 
 approximate agreement~\cite{Attiya1994}, etc, and the ideas at their core can be further developed in order to implement more complex data structures~\cite{Aspnes1990}. Applications 
 of snapshots also appear in sensor networks where snapshot implementations can be used to provide 
 a consistent view of the state of the various sensors of the network.  Under certain circumstances, snapshots 
 can even be used to simulate concurrent graphs, as seen e.g. in~\cite{KK15}. The graph 
 data structure is widely used by many applications, such as the representation of transport networks~\cite{Anez1996}, 
 video-game design~\cite{Bulitko2011}, automated design of digital circuits~\cite{Johannes1996}, making the study 
 of snapshot objects pertinent even to these areas. 

There are many different implementations of snapshot objects based on the progress guarantee that they provide. 
%With the term $progress$, we refer to the ability of an operation to successfully terminate its execution, independently from other operations or failures of other processes in the system. 
However, in order to be fault tolerant against process failure, a concurrent object has to have strong progress guarantees, such as  
{\it wait-freedom}, i.e. the progress guarantee which ensures that an operation invoked by any process that does not fail, returns a result after it executes a finite number of steps.
We provide two wait-free algorithms that implement a snapshot object, namely  
%In this \lena{thesis}{work}, we study the implementations of snapshot objects and provide two algorithms that implement such an object in a wait-free manner.
an algorithm for a \textit{single-scanner} snapshot object, i.e. a snapshot object where only one process is allowed to read the values of the  components, although any process may modify the values of components; and an algorithm for a \textit{$\lambda$-scanner} snapshot object, where up to $\lambda$ predefined processes may read the components of the object, while any process may change the value of any component. Note that $\lambda$ should be lower than or equal to $n$, i.e. the number of processes in the system. In case the value of $\lambda$ is equal to $n$, we obtain a general multi-scanner snapshot object. Our $\lambda$-scanner implementation allows us to study trade-offs, since the increase of the value of $\lambda$ leads to a linear increase of the space and step complexity. 
%
%\textbf{We decided to present a wait-free implementation since it provides a better progress guarantee than other implementations. We also want to explore the lower bound of such algorithms by providing an implementation that uses less space than other state-of-the-art implementations and provides a relatively small step complexity. We also present a trade-off in our $\lambda-scanner\ snapshot$ implementation, since the increase of the $\lambda$ value leads to a linear increase of the space and step complexity.}
%
Our algorithms can be modified to obtain partial snapshot implementations (see Sections~\ref{partial1snap} and~\ref{partialLsnap}), 
where processes execute modified $SCAN$ operations that can obtain the values of just a subset of the snapshot components. 
%We use the term $dynamic$ to refer to the fact that the components, which a $PARTIAL\_SCAN$ reads, can be defined at execution time. Notice that these advantages are important if we use our implementation of the concurrent snapshot object to provide a simulation of a concurrent graph object, such as the graph object presented in \cite{KK15}.

In terms of shared registers, our algorithm $\lambda-Snap$ has a low space complexity of $O(\lambda m)$, where $m$ is the number of the components of the snapshot object. This does not come with major compromises in terms of step complexity, since the step complexity of an $UPDATE$ operation is $O(\lambda )$, while that of a $SCAN$ operation is $O(\lambda m)$. The registers we use are of unbounded size, although the only unbounded value that they store is a sequence number. This is a common practice from many state-of-the-art implementations~\cite{FK17j,Riany2001}. The atomic primitive the registers need to support is $CAS$~\textit{(Compare\ And\ Swap)}, although we present a version of the algorithm using $LL/SC$ registers in order to be more comprehensive and easier to prove correct. An $LL/SC$ register can be constructed by $CAS$ registers using known constructions~\cite{Jayanti2006,Michael2004}. 
%In the Appendix, we present correctness proofs for our algorithms.% $1-Snap$ and $\lambda-Snap$.

The rest of this work is organized as follows. Section~\ref{related} provides a brief comparison of our work with other state-of-the-art algorithms that solve similar problems. Section~\ref{model} exposes the theoretical framework we use. Section~\ref{1snap} presents $1-Snap$, our wait-free implementation of a single-scanner snapshot object, and Section~\ref{lambdasnap} presents $\lambda-Snap$, our wait-free $\lambda$-scanner implementation. %Section~\ref{discussion} is a concluding discussion that may lead to further study in this field.
Section~\ref{discussion} contains a concluding discussion.

\subsection{Related work}
\label{related}
Most of current multi-scanner snapshot implementations that use registers of relatively small size either have step complexity that is linear to the number of processes $n$~\cite{AHR95, Imbs2012} or the space complexity is linear to the number of $n$~\cite{Attiya2008,Imbs2012,Jayanti2002,Jayanti2005,KK15}. The only exception is the multi-scanner snapshot implementation presented by Fatourou and Kallimanis in~\cite{FK07}. However, this snapshot implementation uses unrealistically large registers, since it requires registers that contain a vector of $m$ values as well as a sequence number.
The step complexity of $\lambda-Snap$ is $O(\lambda m)$ for $SCAN$ and $O(\lambda)$ for $UPDATE$, while it uses $O(\lambda m)$ $LL/SC$ registers. In cases where $\lambda$ is a relatively small constant, the number of registers used can be reduced almost to $O(m)$, while the step complexity of $SCAN$ is almost linear to $m$ and the step complexity of $UPDATE$ is almost constant.
Compared to current single-scanner snapshot implementations~\cite{FK06,FK07,FK17j,Jayanti2005,Kirousis1994,Riany2001}, $\lambda-Snap$ %gives a snapshot object 
offers the capability to have more than one $SCAN$ operation at each point of time by slightly worsening the step complexity.
In the worst case where the value of $\lambda $ is equal to $n$, $\lambda -Snap$ provides an implementation of a multi-scanner snapshot object that uses a smaller amount of registers compared to the implementations in~\cite{AHR95,Jayanti2002,Jayanti2005,Riany2001}.
To the best of our knowledge, $\lambda -Snap$ provides the first trade-off between the number of active scanners and the step/space complexity. 

We now compare $\lambda-Snap$ snapshot with other multi-scanner algorithms. 
In Table~\ref{table:related:multi-scanners}, we present the basic characteristics of each snapshot implementation that is reviewed in this section.
Riany et al. have presented in~\cite{Riany2001} an implementation of snapshot objects that uses $O(n^2)$ registers and achieves $O(n)$ and $O(1)$ step complexity for $SCAN$ and $UPDATE$ operations respectively. 
%This implementation uses more registers if $n^2$ is greater than $m$.
Attiya, Herlihy \& Rachman present in~\cite{AHR95} a snapshot object that has $O(n\log^2{n})$ step complexity for both $SCAN$ and $UPDATE$ operations, while it uses dynamic Test\&Set registers. 

Fatourou and Kallimanis~\cite{FK07} present a multi-scanner implementation with $O(m)$ for step complexity $SCAN$ operations and $O(1)$ step complexity for $UPDATE$ operations. In contrast to $\lambda-Snap$, this snapshot implementation requires registers that contain a vector of $m$ values as well as a sequence number. Moreover, the multi-scanner snapshot implementation of \cite{FK07} does not support partial snapshots.

Kallimanis and Kanellou \cite{KK15} present a wait-free implementation of a graph object. This implementation can be slightly modified to simulate a snapshot object, which supports partial $SCAN$ operations. This algorithm manages to implement $UPDATE$ and $SCAN$ operations with step complexity of $O(k)$, where $k$ is the number of active processes in a given execution. It also maintains a low space complexity of $O(n+m)$ but the registers used are of unbounded size. In essence, the algorithm needs registers that can contain $O(n)$ integer values, where half of those values are unbounded.

Imbs and Raynal~\cite{Imbs2012} provide two implementations of a partial snapshot object. The first implementation uses simpler registers than the registers used in the second implementation, but it has a higher space complexity. Thus, we concentrate on the second implementation that achieves a step complexity of $O(nr)$ for $SCAN$ and $O(r_in)$ for $UPDATE$, where $r_i$ is a value that is relative to the helping mechanism the $UPDATE$ operations provide. This implementation uses $O(n)$ Read/Write (abbr. $RW$) and $LL/SC$ registers. Finally, the implementation of Imbs and Raynal provides a new helping mechanism by implementing the ``write first, help later'' technique in their work.

Attiya, Guerraoui and Ruppert \cite{Attiya2008} provide a partial snapshot algorithm that uses $O(m+n)$ CAS registers. The $UPDATE$ operations of this implementation have a step complexity of $O(r^2)$. The step complexity of $SCAN$ is $O({\overline{C}}^2_Sr^2_{max})$, where ${\overline{C}}_S$ is the number of active $SCAN$ operations, whose execution interval overlaps with the execution interval of $S$, and $r_{max}$ is the maximum number of components that any $SCAN$ operation may read in any given execution.

\begin{table}
\footnotesize
\begin{tabular}{p{1.2in}cp{1.3in}p{0.8in}p{0.6in}p{0.9in}}\hline
Implementation                                & Partial      & Regs type             & Regs number     & $SCAN$         & $UPDATE$            \\ \hline
$\lambda$-Snap                                &              & LL/SC \& $RW$          & $O(\lambda m)$  & $O(\lambda m)$ & $O(\lambda )$       \\
partial $\lambda$-Snap                        & $\checkmark$ & LL/SC \& $RW$          & $O(\lambda m)$  & $O(\lambda r)$ & $O(\lambda )$       \\
Attiya,~et.~al.~\cite{AHR95}                  &              & dynamic Test\&Set     & unbounded       & $O(n\log^2{n})$& $O(n\log^2{n})$     \\
Fatourou \&\n Kallimanis~\cite{FK07}          &              & CAS \& $RW$            & $O(m)$          & $O(m)$         & $O(1)$              \\
Jayanti  \cite{Jayanti2005}                   &              & CAS or LL/SC   \& $RW$ & $O(mn^2)$       & $O(m)$         & $O(1)$              \\
Jayanti \cite{Jayanti2002}                    &              & CAS or LL/SC   \& $RW$ & $O(mn^2)$       & $O(m)$         & $O(m)$              \\
Riany\ et\ al. \cite{Riany2001}               &              & CAS or LL/SC \& Fetch\&Inc \& $RW$& $O(n^2)$&$O(n)$       & $(1)$               \\
Kallimanis \&\n Kanellou \cite{KK15}          & $\checkmark$ & CAS or LL/SC   \& $RW$ & $O(n+m)$        & $O(k)$         & $O(k)$              \\
D.\ Imbs\ \&\n M.\ Raynal \cite{Imbs2012}     & $\checkmark$ & LL/SC \& $RW$          & $O(n)$          & $O(nr)$        & $O(r_in)$           \\
Attiya, Guerraoui \& Ruppert \cite{Attiya2008}& $\checkmark$ & CAS  \& $RW$           & $O(n+m)$        & $O(r^2)$       & $O({(\overline{C}_S})^2r^2_{max})$ \\ \hline 
\end{tabular}
\caption{Known multi-scanner snapshot implementations}
\label{table:related:multi-scanners}
\end{table} 

We now compare $\lambda-Snap$ and $1-Snap$ snapshot with other single-scanner algorithms. Recall that $\lambda-Snap$ gives the ability to a snapshot object to have more than one $SCAN$ operation at each point of time by slightly worsening the step complexity. In Table~\ref{table:related:single-scanner}, we present the basic characteristics of each snapshot implementation that is reviewed in this section. 

In~\cite{FK07,FK17j}, Fatourou and Kallimanis provide a single-scanner snapshot implementation, which is called $T-Op$, that achieves $O(1)$ step complexity for $UPDATE$ and $O(m)$ for $SCAN$. By applying some trivial modifications to $T-Opt$, a partial snapshot implementation with $O(r)$ step complexity for $SCAN$ and $O(1)$ for $UPDATE$ could be derived. In contrast to $1-Snap$, $T-Opt$ uses an unbounded number of registers. Moreover, the $RT$ and $RT-Opt$ snapshot implementations presented in~\cite{FK07,FK17j} do not support partial $SCAN$ operations. In~\cite{Jayanti2005}, Jayanti presents a single-scanner snapshot algorithm with $O(1)$ step complexity for $UPDATE$ and $O(m)$ for $SCAN$, while it uses $O(m)$ LL/SC \& $RW$ registers. The algorithm of \cite{Jayanti2005} could be easily modified to support partial $SCAN$ operations without having any negative impact on step and space complexity. Therefore, $1-Snap$ and $\lambda-Snap$ (for $\lambda = 1$) match the step complexity of implementations presented in~\cite{FK07,FK17j,Jayanti2005}, which is $O(m)$ for $SCAN$ and $O(1)$ for $UPDATE$. Denote that the single-scanner implementations of \cite{FK17j,FK07} use $RW$ registers, while $1-Snap$ and $\lambda-Snap$ use $LL/SC$ registers. The partial versions of $1-Snap$ and %the partial version of 
$\lambda-Snap$ (for $\lambda = 1$) have step complexity of $SCAN$ that is reduced to $O(r)$, where $r$ is the amount of components the $SCAN$ operation wants to read.

Kirousis et al. \cite{Kirousis1994} provide a single scanner implementation that uses an unbounded number of registers and has unbounded time complexity for SCAN. A register recycling technique is applied to this snapshot implementation resulting a snapshot implementation with $O(mn)$ step complexity for $SCAN$ and $O(1)$ for $UPDATE$. Riany, et al. \cite{Riany2001} present an algorithm a single-scanner implementation, which is a simplified variant of the algorithm presented in~\cite{Kirousis1994}. This snapshot implementation achieves $O(1)$ step complexity for $UPDATE$ and $O(n)$ for $SCAN$. By applying some trivial modifications, a partial snapshot implementation could be derived. However, the snapshot implementation of~\cite{Riany2001} is a single-updater snapshot object, since it does not allow more than one processes to update the same component at each point of time.

\begin{table}
	\centering
	\footnotesize
	\begin{tabular}{lccccc}\hline
		Implementation                       & Partial      & Regs type        & Regs number & $SCAN$   & $UPDATE$       \\ \hline
		$1-Snap$                             &              & LL/SC \& SW $RW$  & $O(m)$      & $O(m)$   & $O(1)$         \\
		$1-Snap\ (partial)$                  & $\checkmark$ & LL/SC \& SW $RW$  & $O(m)$      & $O(r)$   & $O(1)$         \\
		$Checkmarking$ \cite{FK06,FK17j}     &              &    $RW$           & $m+1$       & $O(m^2)$ & $O(m^2)$       \\
		$T-Opt$~\cite{FK07,FK17j}(modified)  & $\checkmark$ &    $RW$           & Unbounded   & $O(m)$   & $O(1)$         \\
		$RT$ \cite{FK07,FK17j}               &              &    $RW$           & $O(mn)$     & $O(n)$   & $O(1)$         \\
		$RT-Opt$~\cite{FK07,FK17j}           &              &    $RW$           & $O(mn)$     & $O(m)$   & $O(1)$         \\
		$Kirousis~et~al.$ \cite{Kirousis1994}&              &    $RW$           & $O(mn)$     & $O(mn)$  & $O(1)$         \\
		$Riany\ et\ al.$ \cite{Riany2001}    & $\checkmark$ &    $RW$           & $n+1$       & $O(n)$   & $O(1)$         \\
%		$Jayanti$ \cite{Jayanti2005}         &              &    $RW$           & $O(n)$      & $O(n)$   & $O(1)$         \\
		$Jayanti$ \cite{Jayanti2005}         & $\checkmark$ & LL/SC \& $RW$     & $O(m)$      & $O(m)$   & $O(1)$         \\ \hline
	\end{tabular} 
	\caption{Known single-scanner snapshot implementations}
	\label{table:related:single-scanner}
\end{table} 

In~\cite{FK06,FK17j}, Fatourou and Kallimanis provide the $Checkmarking$ algorithm that achieves $O(m^2)$ step complexity for both $SCAN$ and $UPDATE$, while it uses $O(m)$ $RW$ registers. This implementation does not support partial $SCAN$ operations.

\section{Model}
\label{model}
 We consider a system consisting of $n$ uniquely distinguishable processes modeled as sequential state machines, 
 where processes may fail by crashing. 
% A process $p$ may fail by crashing. 
 The processes are asynchronous and communicate through shared  $base\ objects$. A base object stores a value, 
 and it provides a set of $primitives$, through which the object's value can be accessed and/or modified.
\begin{compactenum}
%\item  A $Single-Write\ Multi-Read\ register\ R$ of a process $p$ is a shared object that stores a value from a set and supports the primitives:
%\begin{enumerate}
%\item $Write(R,v)$ that writes the value $v$ in $R$, and can only be invoked by process $p$. This primitive returns a positive acknowledgment $ack$.
%\item  $Read(R)$ that returns the value of $R$, and can be invoked by any process.
%\end{enumerate}
%\item  A $Multi-Write\ Multi-Read\ register\ R$ or more simple $Multi\ Read/Write$ $register\ R$, is a shared object that stores a value from a set and that supports the primitives:
\item  A $Read-Write\ register\ R$ ($RW register$), is a shared object that stores a value from a set and that supports the primitives:
(i) $Write\left(R,v\right)\ $ that writes the value $v$ in $R$, and returns $true$, and 
(ii) $Read(R)$ that returns the value stored in $R$.
\item  An $LL/SC$ $register\ R$ is a shared object that stores a value from a set and supports the primitives:
 (i) $LL(R)$ which returns the value of $R$, and 
(ii) $SC(R,v)$ which can be executed by a process $p$ only after the execution of an $LL(R)$ by the same process. An $SC(R,v)$ writes the value $v$ in $R$ only if the state of $R$ hasn't changed since $p$ executed the last $LL(R)$, in which case the operation returns $true$; it returns $false$ otherwise.
\item  An $LL/SC-Write\ register\ R$ is a shared object that stores a value from a set. It supports the same primitives as an $LL/SC$ $register\ R$ and in addition, the primitive $Write(R,v)$ that writes the value $v$ in $R$, and returns $true$.
\end{compactenum}

 A $shared\ object$ is a data structure that can be accessed and/or modified by processes in the system. Each shared object provides a set of $operations$. Any process can access and/or modify the shared object by invoking operations that are supported by it.
 An $implementation$ of a shared object uses base objects to store the state of the shared object and provides a set of algorithms that use the base objects to implement each operation of the shared object. An operation consists of an $invocation$ by some process and terminates by returning a $response$ to the process that invoked it.
 Similar to each base object, each process also has an internal state. A $configuration\ C$ of the system is a vector that contains the state of each of the $n$ processes and the value of each of the base objects at some point in time. %In essence, a configuration describes the state of the system at some point in time. 
 In an \textit{initial configuration}, the processes are in an \textit{initial state} and the base objects hold an \textit{initial value}. We denote an initial configuration by $C_0$.
 A $step$ taken by a process consists either of a primitive to some base object or the response to that primitive. Operation invocations and responses are also considered steps. Each step is executed atomically.

 An $execution$ $a$ is a (possibly infinite) sequence $C_o,e_1,C_1,e_2,C_2\dots $ , alternating between configurations and steps, starting from some initial configuration $C_o$, where each $C_k,\ k>0$, results from applying step $e_k$ to configuration $C_{k-1}$. If $C$ is a configuration that is present in $a$ we write $C\in a$. An $execution\ interval$ of a given execution $a$ is a subsequence of $a$ which starts with some configuration $C_k$ and ends with some configuration $C_l$ (where $0\le k<l)$. An $execution\ interval$ of an operation $op$ is an execution interval with its first configuration being the one right after the step where $op$ was invoked and last being the one right after the step where $op$ responded.

 Given an execution $a$, we say that a configuration $C_k$ $precedes$ $C_l$ if $k<l$. Similarly, we say that step $e_k$ precedes step $e_l$ if $k<l$. We say that a configuration $C_k$ precedes the step $e_l$ in $a$, if $k<l$. On the other hand, we say that the step $e_l$ precedes $C_k$ in $a$ if $l\le k$.  We furthermore say that $op$ precedes $op'$ if the step where $op$ responds precedes the step where $op'$ is invoked. Given two execution intervals $I,I'$ of $a$, we say that $I$ precedes $I'$ if any configuration $C$ contained in $I$ precedes any configuration $C'$ contained in $I'$.

% Given an execution $a$, we say that operation $op$ is active at a configuration $C$ if $C$ is contained inside the execution interval of $op$, otherwise, $op$ is inactive at $C$.
 An operation $op$ is called $concurrent$ with an operation $op'$ in execution $a$ if there is at least one configuration $C\in a$, such that both $op$ and $op'$ are active in $C$. An execution $a$ is called $sequential$ if in any given $C\in a$ there is at most one active $op$. An execution $a$ that is not $sequential$ is called $concurrent$.
 Executions $a$ and $a'$ are equivalent if they contain the same operations and only those operations are invoked in both of them by the same process, which in turn have the same responses in $a$ and $a'$.

 An execution $a$ is $linearizable$ if it is possible to assign a linearization point, inside the execution interval of each operation $op$ in $a$, so that the response of $op$ in $a$ is the same as its response would be in the equivalent sequential execution that would result from performing the operations in $a$ sequentially, following the order of their linearization points. An implementation of a shared object is linearizable if all executions it produces are linearizable.
%
% A process $p$ may fail by crashing. 
 %In this case, there is a configuration $C\in a$, such that there is no step $e\in a$ that follows $C$ and is executed by $p$. To put it simply, $p$ takes no more steps after a certain point in time. Sometimes we may abuse notation and say that an operation $op$ crashed, meaning that the process $p$ executing it failed by crashing while executing $op$.
%
An implementation $IM$ of a shared object $O$ is $wait-free$ if any operation $op$, of a process that does not crash in $a$, responds after a finite amount of steps. The maximum number of those steps is called $step\ complexity$ of $op$.

 A $snapshot\ S$ is a shared object that consists of $m$ components, each taking values from a set, that provides the following two primitives:
(i) $SCAN()$ which returns a vector of size $m$, containing the values of $m$ components of the object, and (ii) $UPDATE(i,v)$ which writes the non $NULL$ value $v$ on the $i-th$ component of the object.
 A $partial\ snapshot\ S$ is a shared object that consists of $m$ distinct components denoted by $c_o,c_1,\dots ,c_{m-1}$, each taking values from a set, that provides the following two primitives: (i) $PARTIAL\_SCAN(A)$ which, given a set $A$ that contains integer values ranging from $0$ to $m-1$, returns for each $i\in A$ the value of the component $c_i$, and (ii) $UPDATE(i,v)$ which writes the non $NULL$ value $v$ on $c_i$.
A snapshot implementation is $single-scanner$ if in any execution $a$ produced by the implementation there is no $C\in a$ in which there are more than one active $SCAN$ operations. Similarly, a snapshot implementation is $\lambda -scanner$ if in any execution $a$ produced by the implementation there is no $C\in a$ in which there are more than $\lambda $ active $SCAN$ operations.

\section{1-Snap}
\label{1snap}
%\subsection{Description of 1-Snap}
In this section, we present the $1-Snap$ snapshot object (see Listings~\ref{alg:1opt_ds}-\ref{alg:1opt_apply_update}). 

In $1-Snap$, only a single, predefined process is allowed to invoke $SCAN$ operations, while all processes can invoke 
$UPDATE$ operations on any component of the snapshot object. $1-Snap$ uses shared integer variable $seq$, with initial 
value $0$, in order to provide sequence numbers to operations. Each applied operation gets a {\it sequence number} by 
reading the value of $seq$. %We refer to that as the {\it sequence number} of the operation. 
An operation $op$ that is applied with a smaller 
sequence number than that of another operation $op'$ is considered to be applied before $op'$. Since only $SCAN$ operations 
can increase the value of $seq$ by one and since in any given configuration there is only one active $SCAN$ operation 
in our implementation, the $seq$ register can safely be a $RW$ register.
% The data structures of the algorithm are shown in Listing~\ref{alg:1opt_ds}. $1-Snap$ employs the shared $Read-Write$ register $seq$, which takes integer values and its initial value is 0. Each operation reads $seq$ in order to be assigned a sequence number. Only $SCAN$ operations can increase the value of $seq$ by one. Since in any given configuration there is only one active $SCAN$ operation in our implementation, this register can safely be a $RW$ register.

\begin{lstlisting} [float=!t,numbers=left,basicstyle=\scriptsize,language=c,breaklines=true,tabsize=2,numberblanklines=false,countblanklines=false,caption={Data Structures of 1-Snap.},label={alg:1opt_ds},escapechar=@,name=1snap]
struct value_struct {
	val	value;									
	int seq;
	val proposed_value;
};

struct pre_value_struct {
	val value;
	int seq;
};

shared int seq;
shared value_struct values[0..m-1]=[<NULL,NULL,NULL>,...,<NULL,NULL,NULL>]; 
shared pre_value_struct pre_values[0..m-1]=[<NULL,NULL>,...,<NULL,NULL>];

private int view[0..m-1]=[NULL,NULL,...,NULL,NULL];
\end{lstlisting}

$1-Snap$ uses shared vector $values$, consisting of $m$ structs, to represent the components of the snapshot object. 
% $1-Snap$ uses a shared table called $values$ consisting of $m$ structs. 
Each struct of $values$ is stored in an $LL/SC$ register and any process can execute $LL$ and $SC$ operations on each of them. The $i-th$ component of the snapshot object is stored in the $i-th$ struct of the $values$ data structure, this struct is denoted $values[i]$ and its type is $value\_struct$. Each of those structs contains the following three fields:
(1) a $val$ variable called $value$ which stores the value of the $i-th$ component of the snapshot object that is simulated by $1-Snap$,
%The value of a component takes integer values, 
(2) an integer variable called $seq$, which stores the sequence number of the last $UPDATE$ operation that has been applied to the $i-th$ component of the snapshot. This is also referred to as the sequence number of the $i-th$ component, and (3) a val variable called $proposed\_value$ which stores the value that the announced $UPDATE$ operation wants to apply on the $i-th$ component of the snapshot.

This means that each component of the snapshot 
object can store two values, namely its current value and the {\it proposed value}. 
%which is the value that an $UPDATE$ operation attempts to write on this component. 
The process that executes the $SCAN$ operations uses a unique vector $pre\_values$, which consists of $m$ structs that are stored in 
an $LL/SC$ register and any process can execute $LL$ and $SC$ operations on them. The $i-th$ struct of $pre\_values$, 
$pre\_values[i]$, contains a previous value of the $i-th$ component and a sequence number 
of the component of the snapshot object. This sequence number is always smaller than that of the $SCAN$ executed by process $p$. 
Since we apply a helping mechanism, any $UPDATE$ and $SCAN$ operation 
can read and modify the components of this data structure regardless of their process id.
%The process that performs $SCAN$ operations uses a shared table called $pre\_values$. This table consists of $m$ structs that are stored in an $LL/SC$ register and any process can execute $LL$ and $SC$ operations on them. The $i-th$ struct of $pre\_values$ table is denoted by $pre\_values[i]$ and it contains a previous value of the $i-th$ component. In other words, it contains the most recent value of the $i-th$ component of the snapshot object that has a sequence number smaller than that of the last invoked $SCAN$ operation. The $pre\_values[i]$ is a struct of type $pre\_value\_struct$ and it contains the following two variables:
%\begin{enumerate}
%	\item  A val variable called $value$, which stores a copy of the value of the $i-th$ component of the snapshot.
%	\item  An int variable called $seq$, which stores the sequence number of the corresponding component. This sequence number is always smaller than that of the $SCAN$ executed by process $p$.
%\end{enumerate}

A $SCAN$ operation increases the value of $seq$ by one and uses this increased value as its sequence number (line~\ref{seqinc}). 
$UPDATE$ operations that have been applied with a greater or equal sequence number than that of this $SCAN$, 
are not ``visible'' by it (recall that operations are considered to be applied in increasing order %based on 
of their assigned sequence number). Afterwards, for each component of the snapshot object (lines~\ref{for1snap}-\ref{endfor1snap}), the $SCAN$ performs 
the following steps: (1) It tries to copy the value of this component to $pre\_values$ data structure if the sequence number of the component is lower than the sequence number of the corresponding $SCAN$ (lines~\ref{scanitem1}~-~\ref{scanitem1end}). 
(2) It tries to apply an announced $UPDATE$ to this component of the snapshot object (lines~\ref{scanitem2}~-~\ref{scanitem2end}).
(3) Finally, $SCAN$ returns its copy of the snapshot object (line~\ref{1snapreturn}).

An $UPDATE$ operation $U$ on the $j-th$ component executed by process $p$ first tries to announce 
the new value that it wants to store on the $j-th$ component of the snapshot. This is achieved by 
trying to write on the $proposed\_value$ field of the $j-th$ component (lines $18-22$). Afterwards, 
$U$ tries to copy the value of the $j-th$ component of the snapshot to $pre\_values$ data structure if needed (lines 52~-~60). 
Then it tries to update the value of the $j-th$ component of the snapshot using a local copy of $seq$ 
as its sequence number (lines 61~-~67). If the announcement was successful, then the $UPDATE$ operation ends its 
execution after the aforementioned last step. Otherwise, it repeats all previous steps for one last time. 
Doing so will make sure that an $UPDATE$ operation (may or may not be the same as $U$) on the $j-th$ 
component of the snapshot object is applied and furthermore linearized inside the execution interval of $U$.

%A $SCAN$ operation increases the value of $seq$ by one and uses this increased value as its sequence number. $UPDATE$ operations that have been applied with a greater or equal sequence number than that of this $SCAN$, are not ``visible'' by it (recall that operations are considered to be applied in increasing order based on their assigned sequence number). Afterwards, for each component of the snapshot object, the $SCAN$ performs the following steps:
%\begin{enumerate}
%\item  It tries to copy the value of this component to $pre\_values$ data structure if the sequence number of the component is lower than the sequence number of the corresponding $SCAN$. \here{corresponding SCAN?!}
%\item  It tries to apply an announced $UPDATE$ to this component of the snapshot object.
%\end{enumerate}
%Finally, $SCAN$ returns its copy of the snapshot object.
%\here{what to do about line number references?}

\begin{lstlisting} [float=!t,numbers=left,basicstyle=\scriptsize,language=c,breaklines=true,tabsize=2,numberblanklines=false,countblanklines=false,caption={$UPDATE$ and $SCAN$ implementations of 1-Snap.},label={alg:1opt},escapechar=@,name=1snap]
void UPDATE(int j, val value){  
	int i;   
	struct value_struct up_value, cur_value;   
	for (i=0; i<2; i++){     
		cur_value=LL(values[j]);    
		up_value=cur_value;    
		up_value.proposed_value=value;    
		if (cur_value.proposed_value==NULL){     
			if (SC(values[j],up_value)){      
				ApplyUpdate(j);      
				break;     
			}
		}
		ApplyUpdate(j);   
	}  
} 
   
pointer SCAN(){   
	int j;   
	struct value_struct v1;   
	struct pre_value_struct v2;   

	seq=seq+1;    @\label{seqinc}@
	for (j=0;j<m;j++){     @\label{for1snap}@
		ApplyUpdate(j);    
		v1=values[j];    
		v2=pre_values[j];    
		if (v1.seq<seq){     
			view[j]=v1.value;    
		}else{     
			view[j]=v2.value;    
		}   @\label{for1snapend}@
	}    @\label{endfor1snap}@
	return view[0..m-1];  @\label{1snapreturn}@
}
\end{lstlisting}

% The data structures of the algorithm are shown in Listing~\ref{alg:1opt_ds}. $1-Snap$ employs the shared $Read-Write$ register $seq$, which takes integer values and its initial value is 0. Each operation reads $seq$ in order to be assigned a sequence number. Only $SCAN$ operations can increase the value of $seq$ by one. Since in any given configuration there is only one active $SCAN$ operation in our implementation, this register can safely be a $RW$ register.

% $1-Snap$ uses a shared table called $values$ consisting of $m$ structs. Each struct of $values$ is stored in an $LL/SC$ register and any process can execute $LL$ and $SC$ operations on each of them. The $i-th$ component of the snapshot object is stored in the $i-th$ struct of the $values$ data structure, this struct is denoted $values[i]$ and its type is $value\_struct$. Each of those structs contains the following three values:
%(1) a $val$ variable called $value$ which stores the value of the $i-th$ component of the snapshot object that is simulated by $1-Snap$. The value of a component takes integer values (2) an int variable called $seq$, which stores the sequence number of the last $UPDATE$ operation that has been applied to the $i-th$ component of the snapshot. This is also referred to as the sequence number of the $i-th$ component (3) a val variable called $proposed\_value$ which stores the value that the announced $UPDATE$ operation wants to apply on the $i-th$ component of the snapshot.

\begin{lstlisting} [float=!t,numbers=left,basicstyle=\scriptsize,language=c,breaklines=true,tabsize=2,numberblanklines=false,countblanklines=false,caption={$ApplyUpdate$ implementation of 1-Snap.},label={alg:1opt_apply_update},escapechar=@,name=1snap]
 void ApplyUpdate(int j) {
	struct value_struct cur_value;
	struct pre_value_struct cur_pre_value,proposed_pre_value;
	cur_value=LL(values[j]);
	cur_seq=seq;

	for (t=0; t<2; t++) {
		cur_pre_value=LL(pre_values[j]); @\label{scanitem1}@
		cur_value=values[j];
		if (cur_value.seq<seq){
			proposed_pre_value.seq=cur_value.seq;
			proposed_pre_value.value=cur_value.value;
			SC(pre_values[j],proposed_pre_value);
		}
	}  @\label{scanitem1end}@
 
	if (cur_value.proposed_value!=NULL) { @\label{scanitem2}@
		cur_value.value=cur_value.proposed_value;
		cur_value.seq=cur_seq;
		cur_value.proposed_value=NULL;
		SC(values[j],cur_value);
	}  @\label{scanitem2end}@
}
\end{lstlisting}

%The process that performs $SCAN$ operations uses a shared table called $pre\_values$. This table consists of $m$ structs that are stored in an $LL/SC$ register and any process can execute $LL$ and $SC$ operations on them. The $i-th$ struct of $pre\_values$ table is denoted by $pre\_values[i]$ and it contains a previous value of the $i-th$ component. In other words, it contains the most recent value of the $i-th$ component of the snapshot object that has a sequence number smaller than that of the last invoked $SCAN$ operation. The $pre\_values[i]$ is a struct of type $pre\_value\_struct$ and it contains the following two variables:
%\begin{enumerate}
%\item  A val variable called $value$, which stores a copy of the value of the $i-th$ component of the snapshot.
%\item  An int variable called $seq$, which stores the sequence number of the corresponding component. This sequence number is always smaller than that of the $SCAN$ executed by process $p$.
%\end{enumerate}

\subsection{A partial version of 1-Snap}
\label{partial1snap}
% We now present a modified version of $1-Snap$ that implements a partial snapshot object. The data structures used in this modified version of $1-Snap$ remain the same and as shown in Algorithm 4. Furthermore, the pseudocode of the $UPDATE$ operation and the $ApplyUpdate$ function remain the same as shown in Algorithms 5 and 7. A new function is introduced called $Read$ (Algorithm 6). This function is invoked by $SCAN$ operations in order to read the values of the snapshot object.
The $1-snap$ snapshot implementation can be trivially modified in order to implement a partial snapshot object (see Listing~\ref{alg:1opt_partial}). 
In order to do that, a new function $Read$ %(Algorithm~\ref{}), 
is introduced. This function is invoked by $PARTIAL\_SCAN$ operations in order to 
read the values of the components indicated by $A$, which a subset of the 
components of the snapshot object.
% A $SCAN$ operation first increases the value of $seq$ shared variable by one and then executes a for loop. For each integer $j$ that is contained in $A$ ($A$ is the set that contains the different components of the graph a $SCAN$ operation wants to read), the $SCAN$ operation tries to help an $UPDATE$ operation that wants to update the value of $c_j$ component by invoking the $ApplyUpdate$ function. Afterwards, it reads the value of $c_j$ by invoking the $Read$ function.
For each component $c_j$ that is contained in $A$, the $PARTIAL\_SCAN$ operation tries to help 
an $UPDATE$ operation that wants to update the value of $c_j$ by invoking the $ApplyUpdate$. 
Afterwards, it reads the value of $c_j$ by invoking the $Read$ function.
% The only modification in this version of $1-Snap$ is that the $SCAN$ operations do not read every component of the snapshot object, they only read the components of set $A$. The sketch of proof of the partial version of $1-Snap$ is outside of the scope of this work.

%%%% Moving this section to the complexity section:
%Both partial $1-Snap$ and non-partial $1-Snap$ provide the same step complexity to 
%$UPDATE$ operations of $O\left(1\right)$ and have the same space complexity of $O(m)$. 
%However, partial $1-Snap$ provides a step complexity of $O(r)$ to $SCAN$ operations, 
%where $r$ is the number of elements contained in $A$. In contrast, the step complexity 
%that non-partial $1-Snap$ provides to $SCAN$ operation is $O(m)$, higher than that of 
%the partial version, since $r\le m$.

\begin{lstlisting} [float=!t,numbers=left,basicstyle=\scriptsize,language=c,breaklines=true,tabsize=2,numberblanklines=false,countblanklines=false,caption={Partial version of 1-Snap.},label={alg:1opt_partial},escapechar=@,name=1snap-partial]
void PARTIAL_SCAN(A){
	seq=seq+1;
	for each j in A{
		ApplyUpdate(j);
		Read(j);
	}
}

val Read(j){
	struct value_struct v1;
	struct pre_value_struct v2;

	v1=values[j];
	v2=pre_values[j];
	if (v1.seq<seq){
		view[j]=v1.value;
	}else{
		view[j]=v2.value;
	}
	return view[j];
}
\end{lstlisting}

\subsection{Step and space complexity of 1-Snap}
The step complexity of any operation of $1-Snap$ is measured by the number of accesses that are executed in shared registers, inside its execution interval.

We start with the worst-case analysis of $ApplyUpdate$.
\begin{compactenum}
\item  In lines 48-51 only an $LL$ operation is performed at line 50 and a read of shared variable $seq$ (line 51).
\item  Lines 52-60 contain a loop that is executed at maximum two times. In each iteration of this loop, there are executed at maximum two $LL/SC$ operations (the $LL$ of line 53 and the $SC$ of line 58) and one read of line 54.
\item  Lines 61-66 contain just a single $SC$ operation (line 65).
\end{compactenum}

Thus, $ApplyUpdate$ executes $O(1)$ shared memory accesses.

We now proceed with the worst-case analysis of the step complexity of any $UPDATE$. The loop of lines 17-28 can be executed two times at maximum and contains an $LL$ (line 18), an $SC$ (line 22) and two invocations of $ApplyUpdate$ (lines 23 and 27). We previously proved that any $ApplyUpdate$ executes $O(1)$ shared memory accesses. It follows that any $UPDATE$ operation executes $O(1)$ shared memory accesses.

Finally, the worst-case analysis of the step complexity of any $SCAN$ is as follows.
\begin{compactenum}
\item  A write operation on the shared value $seq$ is executed on line 34.
\item  Lines 35-44 contain a loop that is executed exactly $m$ times. In each iteration of the loop an invocation of $ApplyUpdate$ is executed (line 36) and two read operations (lines 37 and 38) are performed.
\end{compactenum}

It follows that any $SCAN$ operation executes $O(m)$ shared memory accesses.

%%% moved here from partial snap:
Both partial $1-Snap$ and non-partial $1-Snap$ provide the same step complexity to 
$UPDATE$ operations of $O\left(1\right)$ and have the same space complexity of $O(m)$. 
However, partial $1-Snap$ provides a step complexity of $O(r)$ to $SCAN$ operations, 
where $r$ is the number of elements contained in $A$. In contrast, the step complexity 
that non-partial $1-Snap$ provides to $SCAN$ operation is $O(m)$, higher than that of 
the partial version, since $r\le m$.

The space complexity of $1-Snap$ algorithm is measured through counting the number of shared registers that are needed for its implementation. The implementation of $1-Snap$ deploys three different shared objects:
\begin{compactenum}
\item  A shared integer variable called $seq$ which is stored in a multi-read/write register.
\item  A shared table called $values$ that is consisted of $m$ $LL/SC$ registers.
\item  A shared table called $pre\_values$ that is consisted of $m$ $LL/SC$ registers.
\end{compactenum}

Thus, our implementation deploys $2m\ LL/SC$ unbounded registers and $1\ RW$ register. It follows that the space complexity of our algorithm is $O(m)$.

 The implementation of 1$-Snap$ presented in this work uses $LL/SC$ registers of unbounded size (one sequence number and two integer values). Although registers should be unbounded it can be proven that they need to have a size of $O(log(s))$, where $s$ is the maximum number of $SCANS$ in a given execution. Thus, in executions that the maximum number of $SCAN$ operation is not too big, $1-Snap$ may use bounded registers.

\begin{theorem}
$1-Snap$ is a wait-free linearizable concurrent single-scanner snapshot implementation that uses $O(m)$ registers, and it provides $O(1)$ step complexity to $UPDATE$ operations and $O(m)$ to $SCAN$ operations.
\end{theorem}

\section{$\lambda$-Snap}
\label{lambdasnap}
%\subsection{Description of $\boldsymbol{\lambda}$-Snap}
In this section, we present the $\lambda-Snap$ snapshot object (see Listings~\ref{alg:lsnap_ds}-\ref{alg:lsnap_apply_update}). 

\begin{lstlisting} [float=!b,numbers=left,basicstyle=\scriptsize,firstnumber=1,language=c,breaklines=true,tabsize=2,numberblanklines=false,countblanklines=false,caption={Data structures of $\lambda$-Snap.},label={alg:lsnap_ds},escapechar=@, name=lsnap]
struct value_struct {
	val value;
	val proposed_value;
	int seq;
};

struct pre_value_struct {
	val value;
	int seq;
};

struct scan_struct {
	int seq;
	boolean write_enable;
};

shared int seq;

shared value_struct values[0..m-1]=[<NULL,NULL,NULL>,...,<NULL,NULL,NULL>];

shared pre_value_struct pre_values[0..@$\lambda$@-1][0..m-1]=[<NULL,NULL>,...,<NULL,NULL>];

shared scan_struct s_table[0..@$\lambda$@-1]=[<NULL,0>,<NULL,0>,...,<NULL,0>];

private int view[0..m-1]=[NULL,NULL,...,NULL,NULL]; 
\end{lstlisting}

In $\lambda-Snap$, only a predefined set of $1 \leq \lambda \leq n$ processes are allowed to invoke $SCAN$ operations, while all processes can perform $UPDATE$ operations on any component. Each applied operation gets a sequence number by reading the shared register $seq$. Sequence numbers assigned both to $SCAN$ and $UPDATE$ operations. More specifically, $SCAN$ operations get a sequence number during the beginning of their execution, while $UPDATE$ operations get an actual sequence number at the point they successfully update the component with their value. We often refer to that as the \textit{sequence number} of the operation. A role of the sequence number is that an operation $op$ with a smaller sequence number than that of another operation $op'$ is considered to be applied before $op'$. Also, a sequence number predetermines which $UPDATE$ operations are visible to a $SCAN$ operation. More specifically, $UPDATE$ operations that have been applied with a greater or equal sequence number than that of the sequence number of a $SCAN$ operation, are not visible from this $SCAN$.

\begin{lstlisting} [float=!t,numbers=left,basicstyle=\scriptsize,language=c,breaklines=true,tabsize=2,numberblanklines=false,countblanklines=false,caption={$UPDATE$ and $SCAN$ implementations of $\lambda$-Snap.},label={alg:lsnap},escapechar=@,name=lsnap]
 void UPDATE(int j, val value){
	struct value_struct up_value, cur_value;
 	for (i=0; i<2; i++){
 		cur_value=LL(values[j]);
 		up_value=cur_value;
 		up_value.proposed_value=value;
 		if (cur_value.proposed_value==NULL){
 			if (SC(values[j],up_value)){
 				ApplyUpdate(j);
 				break;
 			}
		}
		ApplyUpdate(j);
	}
}
 
pointer SCAN(){
	s_table[p_id]={1,seq};
	for (i=0;i<3;i++){
		cur_seq=LL(seq);
		for (j=0;j<@$\lambda$@;j++){
			cur_s_table=LL(s_table[j]);
			if(cur_s_table.seq<seq+2 && cur_s_table.write_enable==1){
				cur_s_table.write_enable=0;
				cur_s_table.seq=seq+2;   
				SC(s_table[j],cur_s_table);
		 	}
		}
		SC(seq,cur_seq+1);	
	}
@~@
	for (j=0;j<m;j++){
		ApplyUpdate(j);
		v1=values[j];
		v2=pre_values[p_id][j];
		if (v1.seq<s_table[p_id].seq){
			view[j]=v1.value;
		} else {
			view[j]=v2.value;
		}
	}
	return view[0..m-1];
} 
\end{lstlisting}

%Any process $p$ that is eligible to perform $SCAN$ operations uses a shared table called $values[p].vals$. This table consists of $m$ structs that are stored in an $LL/SC$ register and any process can execute $LL$ and $SC$ operations on them. The $i$-th struct of $pre\_values[p].vals$ table is denoted by $pre\_values[p].vals[i]$ and it contains the value of the $i$-th component, of the snapshot object that is visible by the last $SCAN$ operation invoked by $p$. In other words, it contains the most recent value of the $i$-th component of the snapshot object that has a sequence number smaller than that of the last invoked $SCAN$ operation by $p$. The $pre\_values[p].vals[i]$ is a struct of type $pre\_value\_struct$ and it contains the following fields: (1) the $value$ field that stores a copy of the weight of the $i$-th component of the snapshot, and (2) the $seq$ field that stores the sequence number of the corresponding component. This sequence number is always smaller than that of the $SCAN$ executed by process $p$.

\begin{lstlisting} [float=!t,numbers=left,basicstyle=\scriptsize,language=c,breaklines=true,tabsize=2,numberblanklines=false,countblanklines=false,caption={$ApplyUpdate$ function of $\lambda$-Snap.},label={alg:lsnap_apply_update},escapechar=@,name=lsnap]
void ApplyUpdate(int j) {
	struct value_struct cur_value;
	struct pre_value_struct cur_pre_value,proposed_pre_value;

	cur_value=LL(values[j]);
	cur_seq=seq;
	for (i=0; i<@$\lambda$@; i++) {
		for (t=0; t<2; t++) {
			cur_pre_value=LL(pre_values[i][j]);
			cur_value=values[j];
			if (cur_value.seq<s_table[j].seq){
				proposed_pre_value.seq=cur_value.seq;
				proposed_pre_value.value=cur_value.value;
				SC(pre_values[i][j], proposed_pre_value);
			}
		}
	}

	if (cur_value.proposed_value!=NULL) {
		cur_value.value=cur_value.proposed_value;
		cur_value.seq=cur_seq;
		cur_value.proposed_value=NULL;
		SC(values[j], cur_value);
	}
}
\end{lstlisting}

For assigning sequence numbers to $SCAN$ and $UPDATE$ operations, $\lambda-Snap$ employs a shared $LL/SC$ register $seq$ (line $10$), which takes integer values. Only $SCAN$ operations are able to increase the value of $seq$ by one (lines $36-46$). In contrast to $1-Snap$, $SCAN$ operations in $\lambda-Snap$ get sequence numbers in more complex way (lines $35$-$47$). More specifically, $SCAN$ operations use a consensus-like protocol in order to increase the $seq$ (using $LL/SC$ instructions) and get a new sequence number. In contrast to $1-Snap$, more than one $SCAN$ operations may get the same sequence number. However, for all $SCAN$ operations that get the same sequence number, the following hold: (1) they are performed by different processes, (2) the increment of the $seq$ register using $LL/SC$ instructions takes place insides their execution interval, and (3) all these $SCAN$ operations are eventually linearized at the same point of the increment of register $seq$.
%(see Section~\ref{proof:lsnap} in the appendix for the linearization and the correctness of $\lambda-Snap$). 
Note that an $UPDATE$ operation $U$, which has been applied with a sequence number greater or equal to the sequence number of some $SCAN$ $S$ operation, is not visible to $S$. Since $U$ is not visible to $S$, $U$ is linearized after $S$. In order to ensure that both $LL/SC$ instructions take place in the execution interval of a SCAN operation and try helping themselves and other $SCAN$ operations, the consensus-like protocol is executed $3$ times (lines $36-46$). 

%Furthermore, to assign a sequence number to any $SCAN$ operation, $\lambda -Snap$ uses a shared table of $\lambda $ components called $s\_table$. Each component of the table is a struct of type $scan_struct$ and is stored in an $LL/SC$ $write$ register. The $p-th$ component of the $s\_table$ is denoted by $s\_table[p]$ and contains the sequence number of the last $SCAN$ operation invoked by $p$. More specifically, $s\_table[p]$ contains the following two values: (1) the sequence number $seq$ that contains the sequence number of the $SCAN$ operation invoked by process $p$, and (2) a boolean variable called $write\_enable$ that is $TRUE$ when the sequence number of the corresponding $SCAN$ is not yet been assigned.

Each process $p$ that is able to execute $SCAN$ operations, owns a shared array of $m$ registers, which is called $pre\_values$ (line~$16$). This array of registers stores a previous value and the sequence number for each component that wants to read. As a first step, each $SCAN$ operation tries to increase the value of $seq$ by executing the consensus-like protocol of lines~$36-46$. Afterwards, for each component of the snapshot object a $SCAN$ operation does the following steps: (1) it tries to copy the value of this component to every  $pre\_values[p]$ data structure that is used by $SCAN$ operations (lines~$66-73$ of $ApplyUpdate$), (2) if the sequence number of the component is lower than that of the sequence number of the corresponding $SCAN$ (line~$70$), it tries to apply an announced $UPDATE$ to this component of the snapshot object (line~$73$), and (3) it returns its copy of the snapshot object (line~$59$).

%$\lambda-Snap$ uses a shared table called $values$ consisting of $m$ structs. Each struct of $values$ is stored in an $LL/SC$ register and any process can execute $LL$ and $SC$ operations on each of them. The $i$-th component of the snapshot object is stored in the $i$-th struct of the $values$ data structure, this struct is denoted $values[i]$ and its type is $value_struct$. Each of those structs contains the following fields: (1) the $value$ field that stores the value of the $i$-th component of the snapshot object that is simulated by $\lambda -Snap$, (2) the $seq$ field that stores the sequence number of the last $UPDATE$ operation that has been applied to the $i$-th component of the snapshot, and (3) the $proposed\_value$ field that stores the value that $UPDATE$ operation wants to apply on the $i$-th component of the snapshot.

We now concentrate on describing $UPDATE$ operations. Each component of the snapshot object stores two values. The first one is the current value of the component (i.e. the $value$ field of $value\_struct$ at line $2$) and the second one is the proposed value (i.e. the $proposed\_value$ field of $value\_struct$), simpler said this is the value that an $UPDATE$ currently wants to write on the component. An $UPDATE$ operation $U$ on $j$-th component executed by some process $p$, it first tries to propose the new value that it wants to store on the $j$-th component of the snapshot. This is achieved by trying to write on the $proposed\_value$ of the $j$-th component of the snapshot object (lines~$22-26$). Afterwards, it tries to copy the current value of the $j$-th component of the snapshot to every $pre\_values[p]$ register (one for each scanner) if needed (lines~$66-76$). Then it tries to $UPDATE$ the value of the $j$-th component of the snapshot using a local copy of $seq$ as its sequence number (line~$71$). If the proposal of the new value was successful, then the $UPDATE$ operation ends its execution (line~$26$). Otherwise, it repeats all previous steps for one last time. Doing so will make sure that an $UPDATE$ operation (may or may not be the same as $U$) on the $j$-th component of the snapshot object is applied and furthermore linearized inside the execution interval of $U$. By writing a sequence number with its value, an $UPDATE$ operation $U$ that has been applied with a sequence number less or equal to the sequence number of some $SCAN$ $S$ operation is visible to $S$.

In $\lambda-Snap$, we employ a helping mechanism where $UPDATE$ and $SCAN$ operations try to help $UPDATE$ operations that are slow or stalled (lines $77-82$). More specifically, an $UPDATE$ operation on some component $j$ helps at most $2$ $UPDATE$ operations on the $j$-th component (see lines $77-82$). On the other hand, a $SCAN$ operation helps at most $2$ $UPDATE$ operations per component that it reads. Thus, the non partial version of $\lambda-Snap$ helps at most $2m$ $UPDATE$ operations (in the case of the partial version of $\lambda-Snap$, a $SCAN$ operation helps at most $\lambda r$ $UPDATE$ operations, where $r$ is the number of components that wants to read).

\subsection{A partial version of $\lambda$-Snap}
\label{partialLsnap}

We now present a slightly modified version of $\lambda-Snap$ (see Listing~\ref{alg:lsnap_partial}) that implements a partial snapshot object. The data structures used in this modified version of $\lambda -Snap$ remain exactly the same, as shown in Listing~\ref{alg:lsnap_ds}. Furthermore, the pseudocode of $UPDATE$ and $ApplyUpdate$ function remain the same as shown in Listings~\ref{alg:lsnap} and~\ref{alg:lsnap_apply_update}. A new function is introduced called $Read$ (Listing~\ref{alg:lsnap_partial}). This function is invoked by $PARTIAL_SCAN$ operations in order to read the values of the snapshot object.

\begin{lstlisting} [float=!b,numbers=left,basicstyle=\scriptsize,firstnumber=1,language=c,breaklines=true,countblanklines=false,tabsize=2,numberblanklines=false,caption={$UPDATE$ and $SCAN$ implementations for the partial version of $\lambda$-Snap.},label={alg:lsnap_partial},escapechar=@,name=lsimopt]
pointer	PARTIAL_SCAN(set A) {
	s_table[p_id]={1,seq};
	for	(i=0; i<3; i++) {
		cur_seq=LL(seq);
		for	(j=0;j<@$\lambda$@;j++) {
			cur_s_table=LL(s_table[j]);
			if(cur_s_table.seq<seq+2 && cur_s_table.write_enable==1) {
				cur_s_table.write_enable=0;
				cur_s_table.seq=seq+2;
				SC(s_table[j],cur_s_table);
			}
		}
		SC(seq,cur_seq+1);	
	}

	for	each j in A  {
		ApplyUpdate(j);
		Read(j);
	}
}

val Read(int j){
	struct value_struct v1;
	struct pre_value_struct v2;

	v1=values[j];
	v2=pre_values[j];
	if (v1.seq<seq){
		view[j]=v1.value;
	}else{
		view[j]=v2.value;
	}
	return view[j];
}
\end{lstlisting}

The only modification in this version of $\lambda-Snap$ is that the $PARTIAL\_SCAN$ operations do not read every component of the snapshot object, they only read the components of set $A$. 
%A $PARTIAL\_SCAN$ operation through helping tries to get a sequence number using the consensus-like protocol and then executes a for loop in order to read the components of the snapshot object.
For each component $j$ that is contained in $A$ (the set of components that a $SCAN$ wants to read), the $PARTIAL\_SCAN$ operation tries to help $UPDATE$ operations on the $j$-th component by invoking the $ApplyUpdate$ function (lines~$15-18$). Afterwards, it reads the value of the $j$-th component by invoking the $Read$ function.

Both partial $\lambda -Snap$ and non-partial $\lambda -Snap$ have the same step complexity of $UPDATE$ operations, and the same space complexity. Although, $\lambda -Snap$ provides a step complexity to $SCAN$ operations of $O(\lambda r)$ where, $r$ is the number of components that the $PARTIAL\_SCAN$ operation reads.

\subsection{Step and space complexity of $\lambda$-Snap}
The step complexity of an operation of $\lambda -Snap$ is measured by the number of operations that are executed in shared registers, inside its execution interval.

We start with the worst-case analysis of $ApplyUpdate$.
\begin{enumerate}
\item	In lines 62-65 only an $LL$ operation is performed at line 68 and a read of shared variable $seq$ (line 69).

\item	In lines 66-76 contain a loop that is executed exactly $\lambda $ times. In any iteration of this loop the loop of lines 67-75 is executed exactly two times. In any iteration of the later loop, four shared register operations are executed at maximum. An $LL$ at line 68, two read operations (line 69 and 70) and an $SC$ operation at line 73. Thus, the loop of lines 67-75 executes at maximum eight shared register operations. Furthermore, the loop of lines 67-75 is a nested loop of that of lines 66-76, so it is executed exactly $\lambda $ times. It follows that the loop of lines 66-76 executes at maximum $8\lambda$ shared register operations. 

\item	Lines 77-82 contain just a single $SC$ operation (line 81).
\end{enumerate}

It follows that $ApplyUpdate$ executes at maximum $3+6\lambda $ shared memory accesses. Thus, $ApplyUpdate$ has a step complexity of $O(\lambda $).

We now proceed with the worst-case analysis of the step complexity of any $UPDATE$. The loop of lines 21-32 can be executed two times at maximum and contains an $LL$ (line 22), an $SC$ (line 26) and two invocations of $ApplyUpdate$ (lines 27 and 31). We previously proved that any $ApplyUpdate$ executes $O(\lambda )$ shared memory accesses. It follows that any $UPDATE$ operation executes $O(\lambda )$ shared memory accesses.

We can finally proceed with the worst-case analysis of step complexity of any $SCAN$.
\begin{enumerate}
\item	A write operation on the shared table $s\_table$ is executed in line 35.

\item	Lines 36-47 contain a loop that is executed exactly three times. In each iteration of that loop, an $LL$ is executed at line 37 and an $SC$ at line 46. Furthermore, the loop of lines 38-45 is executed, and exactly $\lambda $ iterations of it are performed. In any iteration of loop of lines 38-45 at maximum three shared memory accesses are performed (an $LL$ at line 39, a read of the shared $seq$ variable at line 40 and an $SC$ at line 43). It follows that the loop of lines 38-45 executes $O(\lambda )$ shared memory accesses. Since the loop of lines 36-47 is executed exactly three times it executes $O\left(\lambda \right)$ shared memory accesses.

\item	Lines 49-58 contain a loop that is executed exactly $m$ times. In each iteration of that loop an $ApplyUpdate$ is invoked (line 50) and two read operations are performed (lines 51, 52). Since $ApplyUpdate$ executes $O(\lambda )$ shared memory accesses and at lines 49-58 is invoked exactly $m$ times it follows that lines 49-58 execute $O(\lambda m)$ shared memory accesses.
\end{enumerate}

\noindent It follows that any $SCAN$ operation executes $O(\lambda m)$ shared memory accesses.

The space complexity of $\lambda -Snap$ algorithm is measured through counting the number of shared registers that are needed for its implementation. The implementation of $\lambda -Snap$ deploys four different shared objects:
\begin{enumerate}
\item	A shared $LL/SC$ register called $seq$.
\item	A shared array called $values$ that is consisted of $m$ $LL/SC$ registers.
\item	A shared array called $pre\_values$ that is consisted of $\lambda m$ $LL/SC$ registers.
\item	A shared array called $s\_table$ that is consisted of $\lambda $ $LL/SC\ write$ registers.
\end{enumerate}

Thus, our implementation deploys $1+m+\lambda m+\lambda \ LL/SC\ write$  registers. It follows that the space complexity of our algorithm is $O(\lambda m)$.

%The implementation of $\lambda -Snap$ presented in this work uses $LL/SC\ write$ registers of unbounded size. Although, registers should be unbounded it can be proven that they need to have a size of $O(log(s))$, where $s$ is the maximum number of $SCANS$ in a given execution. Thus, in executions that the maximum number of $SCAN$ operation is relative low $\lambda-Snap$ may use bounded registers.

\begin{theorem}
$\lambda -Snap$ is a wait-free linearizable concurrent $\lambda$-scanner snapshot implementation that uses $O(\lambda m)$ registers, and it provides $O(\lambda )$ step complexity to $UPDATE$ operations and $O(\lambda m)$ to $SCAN$ operations.
\end{theorem}

\section{Discussion}
\label{discussion}
%In this work an object called $\lambda-Snap\ snapshot$ is proposed. An implementation of such an object provides a solution to the single-scanner snapshot problem and the multi-scanner snapshot problem simultaneously. 
This work proposes the  $\lambda-Snap\ snapshot$ object and its implementations, providing a solution to the single-scanner snapshot problem and the multi-scanner snapshot problem simultaneously. 
If $\lambda $ is equal to 1, then our algorithm simulates a single-scanner snapshot object, while if $\lambda$ is equal to the maximum number of processes,
 then it simulates a multi-scanner snapshot object. %Our $\lambda-Snap\ snapshot$ object provides a new idea to solve the snapshot problem. 
 To the best of our knowledge, there is no publication that provides a solution to the snapshot problem that can support a preset amount of $SCAN$ operation that may run concurrently.

$1-Snap$ solves the single-scanner flavor of snapshot problem. Although, in our algorithm, we only allow one process with a certain id to invoke $SCAN$ operations, this is a restriction that can be easily lifted. The system can support invocations of $SCAN$ operations by any process, although only one process can be active in any given configuration of the execution. In this case, our algorithm would be correct only in executions that no more than one $SCAN$ is active in any given configuration of the execution.

A $\lambda -Snap\ snapshot$ can efficiently applied in systems where only a preset amount of processes may want to execute $SCAN$ operations. Especially in systems that the amount of processes that may want to invoke a $SCAN$ operation is small enough, our algorithm has almost the same performance as a single-scanner snapshot object. An example of such a system may be a sensor network, where many sensors are communicating with a small amount of monitor devices. In this case, sensors essentially perform $UPDATE$ operations while monitor devices may invoke $SCAN$ operations.

\bibliographystyle{plainurl}
\bibliography{paper}

%\clearpage
%\setcounter{page}{0}
%\pagenumbering{arabic}
%\appendix
%
%\section{Appendix: Linearization of 1-Snap}
%\label{proof:1snap}
%\input{1opt_proof}
%
%\section{Appendix: Linearization of $\lambda$-Snap}
%\label{proof:lsnap}
%\input{lopt_proof}

\end{document}